\documentclass[twocolumn,showpacs,preprintnumbers,amsmath,amssymb,showlabels]{revtex4}

\usepackage{graphicx}
\usepackage{dcolumn}
\usepackage{bm}

\begin{document}

\title{Wetting on Random Roughness: the Ubiquity of Wenzel Prewetting}

\author{S. Herminghaus}

\affiliation{Max Planck Institute for Dynamics and Self-Organization (MPIDS), 37077 Goettingen, Germany}

\date{\today}

\begin{abstract}
The wetting properties of solid substrates with  macroscopic  random roughness are considered as a function of the microscopic contact angle of the wetting liquid and its partial pressure in the surrounding gas phase. It is shown that Wenzel prewetting, which has been recently predicted for a rather wide class of roughness profiles derived from Gaussian random processes by a general distortion procedure \cite{EPJE}, should in fact be ubiquitous and prevail under even much milder conditions. The well-known transition occurring at Wenzel's angle is accompanied by a prewetting transition, at which a jump in the adsorbed liquid volume occurs. This should be present on most surfaces bearing homogeneous, isotropic random roughness.
\end{abstract}


\maketitle

While the physics of wetting and spreading on ideally smooth surfaces has meanwhile reached a status of mature textbook knowledge \cite{DeGennes1985,DietrichReview,DeGennesBuch,BonnReview}, the wetting properties of randomly rough solid substrates are still poorly understood. This is in part due to the vast range of scales to be covered, which extends from the nanometer scale of atomic roughness to the scale of millimeters, just before gravity comes into play. There has already been a lot of work concentrating on small scales, where the interplay between interface topography and van der Waals forces play a role \cite{Joanny1984,Jansons1985,AndelRob1988,Netz1997,Chow1998,Swain1998,Seemann2001}, but the  larger scale, ranging from sevelar microns to about a millimeter, which corresponds to typical roughness encountered in practical situations, has been dealt with so far only for a small class of rather special cases. Most authors have tried to model rough surfaces as Gaussian random processes \cite{Nayak1973,Green1984,Ogilvy1989,ColeKrim1989,KardInd1990,PalasKrim1993,Kalli2010,RodValve2010}, or to capture single aspects of wetting using simplified model geometries 
\cite{ChengCole1990,Napiorkowski1993,RasconParry2000,BicoQuere2001,BicoThiele2002,SeeBrinkPNAS,AnnuRev2008,Courbin2009}. 

Since the most common reasons for roughness, like wear and weathering, or even deliberate ones like etching or sand blasting, are the effect of very many more or less independent local attacks, the application of the  Gaussian random model appears at first glance as the most natural, and has consequently been used in a large number of studies
\cite{LongHigg1957,LongHigg1957b,Nayak1973,Green1984,ColeKrim1989,PalasKrim1993}.  However, it has been recently shown that even slight deviations from purely Gaussian  roughness give rise to {\it qualitative} changes in the wetting behaviour of the substrate \cite{EPJE}. Using a simple model of a distorted Gaussian distribution, it was shown that a wetting phase transition appears for a large class of systems which is absent for undistorted Gaussian roughness. This transition, which was termed {\it Wenzel prewetting}, may be of large potential interest in many areas of technology, as well as geosciences and biology. The generality of that study, however, was quite limited, since the underlying random process was still assumed Gaussian, and ample use of multivariate normality had to be made \cite{LongHigg1957,LongHigg1957b,Green1984,Nayak1973}. In the present paper, it is shown that Wenzel prewetting is in fact a {\it universal feature} of randomly rough surfaces, under only very mild conditions which are fulfilled by most customary surfaces.

Wenzel was the first to report on systematic studies of wetting on randomly rough surfaces \cite{Wenzel1936}. He characterized the roughness by a single parameter, ${\rm r}$, which he defined as the ratio of the total substrate area divided by the projected area. Obviously, ${\rm r} \ge 1$, and  ${\rm r}=1$ corresponds to a perfectly smooth surface. The free energy which is gained per unit area when the rough substrate is covered with a liquid is then given by ${\rm r} (\gamma_{sg}-\gamma_{sl})$, where $\gamma_{sl}$ and $\gamma_{sg}$ are the solid-liquid and solid-gas interfacial tension, respectively. If this gain is larger than the surface tension of the liquid, $\gamma$, we expect a vanishing {\it macroscopic} contact angle, because covering the substrate with the liquid releases more energy than is required for the formation of a free liquid surface of the same (projected) area. More specifically, force balance at the three-phase contact line yields
\begin{equation}
\cos\theta_{macro} = \frac{{\rm r} (\gamma_{sg}-\gamma_{sl})}{\gamma} = {\rm r} \cos\theta
\label{Eq:Wenzel}
\end{equation}
where $\theta = \arccos \left[(\gamma_{sg}-\gamma_{sl})/\gamma\right]$ is the microscopic contact angle \cite{Young1805}.  When $\theta$  is reduced to $\theta_W = \arccos (1/{\rm r})$, $\theta_{macro}$ vanishes, and the substrate is covered with an 'infinitely' thick liquid film. In the present paper, we will discuss this transition in some detail, both at liquid-vapour coexistence and below the saturated vapour pressure. 
 
\begin{figure}[h]
\includegraphics[width = 7.5cm]{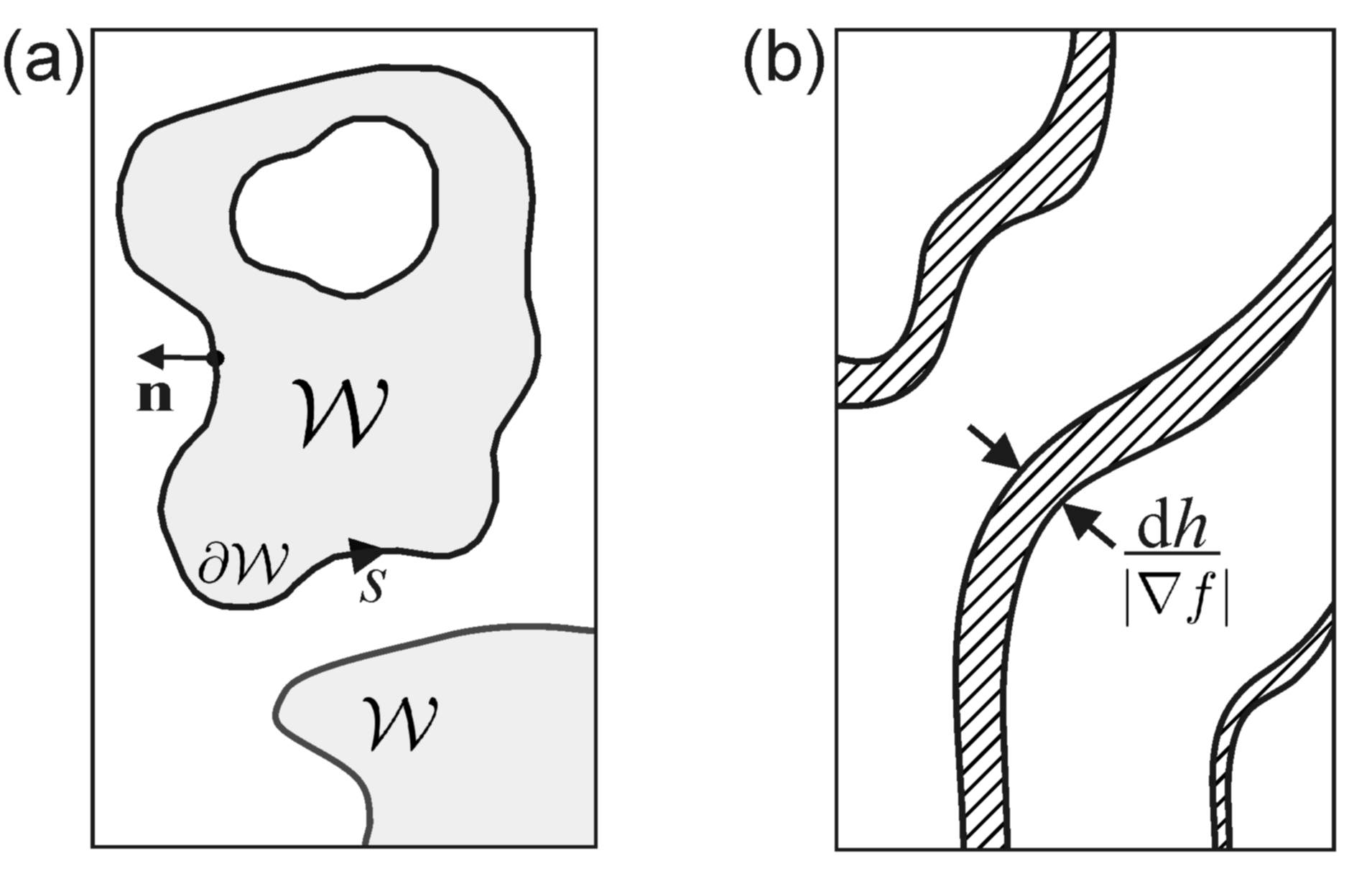}
\caption{(a) Top view of the sample, showing the wetted areas in grey, the bare substrate in white. The normal vector to $\partial\mathcal{W}$, ${\bf n}$, lies in the $(x,y)$-plane. (b) Two sets of contour lines of $f$ at heights $h$ and $h+\ {\rm d}h$. The hatched area between the lines is equal to $p(h) \ {\rm d}h$. \label{Sketch} }
\end{figure}

We describe the rough solid substrate by a function  $f(x,y)$, which shall approximate the actual physical surface at any required precision, but be mathematically smooth, such that $\nabla f$ and $\Delta f$ exist everywhere. The roughness is assumed to be homogeneous and isotropic, i.e., its statistical parameters shall be the same everywhere on the sample, and independent of rotation of the sample about its normal axis. A small amount of  liquid deposited on this substrate will make an interface with the surrounding gas, which is described by a second function, $g(x,y)$. The support of $g$ is the wetted area, which we call $\mathcal{W}$. Continuity of the liquid surface assures $g = f$ on the boundary of $\mathcal{W}$, i.e., the projection of the three-phase contact line, henceforth denoted by $\partial\mathcal{W}$ (cf. Fig.~\ref{Sketch}a).

As the amplitude of most natural roughness is much smaller than its dominant  lateral length scale, we assume that 
\begin{equation}
| \nabla f | \ll 1
\label{Eq:FlatApprox}
\end{equation} 
which allows for substantial simplifications. The same shall hold for $g$. The contact angle with the substrate, $\theta$, yields the boundary condition 
\begin{equation}
| \nabla (g - f)| \ \approx \tan\theta \ \approx \theta
\label{Eq:YoungApprox}
\end{equation}
which is to be fulfilled everywhere on $\partial\mathcal{W}$, to first order in $\theta$, $\nabla f$, and $\nabla g$. Applying Green's theorem to $(g-f)$, we obtain
\begin{equation}
\int_{\partial\mathcal{W}} {\bf n} \cdot \nabla (g-f) \ {\rm d}s= \int_{\mathcal{W}}\Delta (g-f)  \ {\rm d}^2{\bf x}
\label{Eq:GreenWet}
\end{equation}
where $s$ is the distance along $\partial\mathcal{W}$, ${\bf n}$ its unit normal vector, and ${\bf x} = (x,y)$. Since $g = f$ on $\partial\mathcal{W}$, $\nabla (g-f)$ is everywhere perpendicular to $\partial\mathcal{W}$. Hence  eq.~(\ref{Eq:YoungApprox}) may be written as ${\bf n} \cdot \nabla (g-f) \approx \theta$, and eq.~(\ref{Eq:GreenWet}) can be recast into
\begin{equation}
l \  \theta + \int_{\mathcal{W}} [2 H -\Delta f] \ {\rm d}^2{\bf x} = 0 
\label{Eq:Result}
\end{equation}
in which $l$ denotes the length of $\partial\mathcal{W}$, and $H\approx \frac{1}{2}\Delta g$ is the mean curvature of the liquid-vapour interface. The latter is given by the Kelvin equation, 
\begin{equation}
H = \frac{k_B T}{2 \gamma v_m} \ln \frac{{\rm p}_s}{{\rm p}} 
\label{Eq:Kelvin}
\end{equation}
where ${\rm p}$ is the partial pressure of the adsorbed liquid in the surrounding gas phase, ${\rm p} _s$ is its saturated vapor pressure, $v_m$ its molecular volume, and $k_B$ is Boltzmann's constant. It is thus a convenient measure for the deviation from liquid-vapour coexistence.

Before we can exploit eq.~(\ref{Eq:Result}), we derive a few useful relations. Let $p(f)$ is the height distribution of $f({\bf x})$, with normalization $\int p(h) {\rm d}h = |\mathcal{S}|$ (total sample area). Then the total area between the contour lines at $f=h$ and at $f=h+{\rm d}h$ is given by $p(h)\ {\rm d}h$, which corresponds to the hatched area in Fig.~\ref{Sketch}b. The average slope on that set, $\sigma_1(h) =  \langle |\nabla f|\rangle_h$, is given by
\begin{equation}
\sigma_1 = \frac{\int |\nabla f(h)| \ {\rm d}\tau}{p(h) {\rm d}h}
\end{equation}
where ${\rm d}\tau = {\rm d}s{\rm d}h/|\nabla f(h)|$ is the differential of the hatched area. Hence  for the total length of the contour line at height $h$ we obtain 
\begin{equation}
L(h) = \int {\rm d}s = \sigma_1 (h) p(h)
\label{Eq:ContourLength}
\end{equation}
In order to express the integral over $\Delta f$ which appears in eq.~(\ref{Eq:Result}), we apply Green's theorem again, this time to the area enclosed by a contour line. This yields 
\begin{equation}
\int_{\mathcal{C}(h)}\Delta f \ {\rm d}^2{\bf x} = \int_{\partial\mathcal{C}} {\bf n}\cdot \nabla f \ {\rm d}s
\label{Eq:GreenDry}
\end{equation}
where $\mathcal{C}(h)$ denotes the set $\{ {\bf x} \mid f({\bf x})\leq h\}$, and $\partial\mathcal{C}$ its boundary, i.e., the contour line itself. Introducing $\sigma_2 (h) = \langle |\nabla f|^2\rangle_h$, we readily see that eq.~(\ref{Eq:GreenDry}) can be rewritten as 
\begin{equation}
\int_{\mathcal{C}(h)}\Delta f \ {\rm d}^2{\bf x} = \sigma_2 (h) \ p(h)
\label{Eq:GreenDrySigma2}
\end{equation}

In order to fulfill the boundary condition, eq.~(\ref{Eq:YoungApprox}), the vertical position of the three-phase contact line, which may be symbolically written as $f(\partial\mathcal{W})$, will vary along $\partial\mathcal{W}$ about an average value, $\langle f(\partial\mathcal{W})\rangle$. The projection of the contact line onto the plane will thus approximately follow the contour line at $f({\bf x}) = \langle f(\partial\mathcal{W})\rangle$, with excursions towards both the outside and the inside of $\mathcal{W}$. These will in cases represent detours, sometimes shortcuts with respect to $\partial\mathcal{W}$. As a reasonable approximation, we may thus use $l \approx L (\langle f(\partial\mathcal{W})\rangle)$ for the length of the three-phase contact line. Similarly, we set 
\begin{equation}
\int_{\mathcal{W}} \ {\rm d}^2{\bf x}\   \approx \int\limits_{-\infty}^{h}p(f)\ {\rm d}f = W(h) 
\label{Eq:MCApprox}
\end{equation} 
for the wetted sample area, with $h = \langle f(\partial\mathcal{W})\rangle$.  Inserting these expressions in eq.~(\ref{Eq:Result}), we obtain
\begin{equation}
2 H W(h) \approx \left[\frac{\sigma_2(h)}{\sigma_1(h)} -  \theta \right] L(h)
\label{Eq:Adsorb}
\end{equation}
This allows, if $p(h)$, $\sigma_1(h)$, and $\sigma_2(h)$ are known from experimental characterization of the sample, to determine the adsorbed amount of liquid as a function of $\theta$ and $H$. 

Up to this point, we have not made any specific assumption about the roughness profile, except its being sufficiently shallow for the approximations made above to hold. Now we shall go one step further, observing that roughness profiles generated by wear, weathering, erosion, etching, sand blasting, or similar processes will invariably have a finite codomain (we disregard fissures and cracks here). In other words, the support of $p(h)$ is the interval $\left[h_-,h_+\right] \forall {\bf x}$, where $h_-$ represents the depth of the deepest trough, and $h_+$ the height of the highest elevation on the sample.  This has severe consequences of $\sigma_1$ and $\sigma_2$, as both must go to zero as $h\rightarrow h_{\pm}$. To see that, consider the distribution of minima, $\mu_-(h)$. Clearly, $\mu_-\rightarrow 0$ for $h\rightarrow h_-$, and we may choose to write $\mu_-(h)\simeq \mu_0(h-h_-)^{\nu}$, with some $\nu > 0$. Since each minimum is parabolic to first order, one can readily verify that it contributes $4\pi (h-h_0)$ to $\sigma_1$, if $h_0$ is the depth of the minimum. We thus have 
\begin{equation}
\sigma_1(h) \leq 4\pi \int\limits_{h_-}^h \mu_-(z)(h-z) \ {\rm d}z \simeq \frac{\mu_0 (h-h_-)^{\nu+2}}{(\nu+1)(\nu+2)} 
\label{Eq:SigmaMu}
\end{equation}
The '$\leq$' sign has been used because there may be saddle points (or even maxima) occurring at elevations between $h_-$ and $h$, which can only reduce $\sigma_1$. As eq.~( \ref{Eq:SigmaMu}) shows, $\sigma_1(h)$ is bounded from above by a function which vanishes at least quadratically as $h\rightarrow h_-$. An analogous result is obtained for  $h\rightarrow h_+$. 

Next we consider the ratio $\sigma_1/\sigma_2$. With the abbreviation $\rho = | \nabla f |$, we have 
\begin{equation}
\sigma_1 (h) = 2\pi \int \rho^{2} q(h,\rho) {\rm d}\rho
\label{Eq:Sigma1}
\end{equation}
where $q(h,\rho)$ is the distribution of slopes, sampled at height $h$. Similarly, we have 
\begin{equation}
\sigma_2 (h) = 2\pi \int \rho^3 q(h,\rho)\ {\rm d}\rho
\label{Eq:Sigma2}
\end{equation}
As a consequence, 
\begin{equation}
\frac{\sigma_2(h)}{\sigma_1(h)} = \frac{\int \rho^3 q(h,\rho)\ {\rm d}\rho}{\int \rho^2 q(h,\rho){\rm d}\rho} < \rho_{max}(h)
\label{Eq:Sigma2Sigma1}
\end{equation}
where $\rho_{max}$ is the maximum slope encountered at elevation $h$. Since this vanishes as $h\rightarrow h_-$ as $\sigma_1$ does, so will the ratio $\sigma_2/\sigma_1$, as eq.~(\ref{Eq:Sigma2Sigma1}) shows. 

Aside from these global properties, both $\sigma_1$ and $\sigma_2$ are expected to be largely featureless, due to the general fact that the processes leading to roughness exhibit only very limited lateral correlation. For any pronounced feature to develop in $\sigma_i$, distant places on the sample would have to 'conspire' to contribute to that feature at the same depth. This can happen only for composite surfaces, where the roughness topography may penetrate through a coating or other stratigraphic variation of material properties. Such ramifications are interesting to consider for practical purposes, but well beyond the scope of the present article.  

\begin{figure}
\includegraphics[height = 10cm]{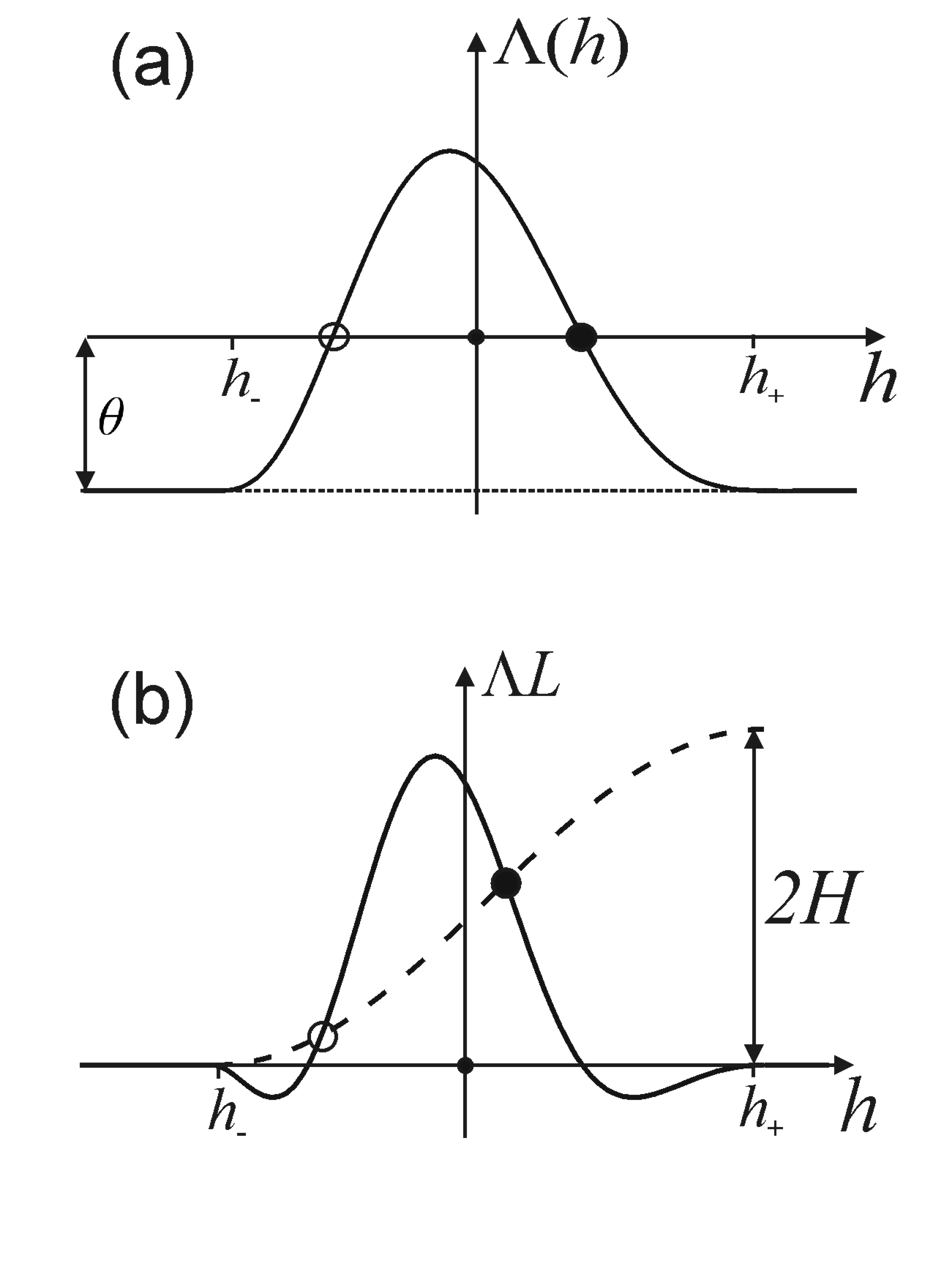}
\caption{Graphic construction for solving eq.~(\ref{Eq:Adsorb}).  The dashed line represents the l.h.s. of   eq.~(\ref{Eq:Adsorb}). \label{LambdaL} }
\end{figure}

The generic shape of the function
\begin{equation}
\Lambda (h) = \left[\frac{\sigma_2(h)}{\sigma_1(h)} -  \theta \right] 
\label{Eq:Lambda}
\end{equation}
which appears in eq.~(\ref{Eq:Adsorb}) is sketched in Fig.~\ref{LambdaL}, according to the discussion above. Following eq.~(\ref{Eq:Adsorb}), the film thickness at coexistence ($H=0$) can be derived from the zeros of $\Lambda$, of which there are either two or none, depending on $\theta$. In the latter case, the contact angle  is too large for forming a liquid surface between the spikes and troughs which complies with the boundary condition, eq.~(\ref{Eq:YoungApprox}). If, however, $\Lambda (h)$ intersects the $h$-axis, the slopes of the zeros decide upon the stability of the corresponding solutions. This can be seen by appreciating that $\Lambda$ may be interpreted as a deviation from the force balance expressed by eq.~(\ref{Eq:YoungApprox}) \cite{Young1805}. For the left zero, which is marked by an open circle in the figure, a displacement of the three-phase contact line would give rise to an imbalance of wetting forces which drives it further away from the zero. The opposite is true for the right zero, marked by the closed circle. The latter therefore corresponds to the stable solution, and thus to the adsorbed film thickness which will develop.

All this is in marked contrast to Gaussian roughness, for which $\sigma_1$ and $\sigma_2$ are independent of $h$, with $\sigma_2/\sigma_1 = 4/\pi \ \forall h$ \cite{EPJE}. This is in fact a dramatic difference, as $\Lambda$ would then just be a horizontal straight line which lies either above or below the $h$ axis depending on $\theta$. As a consequence, the whole structure we are developing here would be absent.

The graph of $\Lambda(h)$ makes a first contact with the horizontal axis when $\theta$ reaches
\begin{equation}
\theta_p = {\rm max}\left(\frac{\sigma_2}{\sigma_1}\right)
\end{equation}
At this point, the formerly dry substrate is covered with a liquid film of 'thickness' $h_p = {\rm argmax}(\sigma_2/\sigma_1)$. A quantity of particular interest is the  total liquid volume, $V$, adsorbed at given $\theta$ and $H$. This is related to $h$ via 
\begin{equation}
V = \int\limits_{h_-}^h(h-f)p(f) {\rm d}f
\label{Eq:Volume}
\end{equation}
and can be evaluated if $h$ and $p(f)$ are known. However, we continue here to discuss $h$ instead, since it is more accessible through the formalism developed above.

It is important to note that $\theta_p$ always lies above $\theta_W$. To see this, we note that ${\rm r} = 1/\cos \theta_W \approx  1+\frac{1}{2}\theta_W^2$. Furthermore, ${\rm r} = \langle \sqrt{1+|\nabla f|^2}\rangle$. Thus we have
\begin{equation}
\theta_W^2 \approx \frac{1}{\mathcal{|S|}} \int\limits_{h_-}^{h_+} \sigma_2(h) p(h){\rm d}h < {\rm max} (\sigma_2)
\label{Eq:WenzelNonGauss}
\end{equation}
On the other hand, $\sigma_2 > \sigma_1^2$, such that 
\begin{equation}
\theta_p^2 = {\rm max}\left(\frac{\sigma_2}{\sigma_1}\right)^2  > {\rm max} (\sigma_2)
\label{Eq:ThetaPrewet}
\end{equation}
From eqs.~(\ref{Eq:WenzelNonGauss}) and (\ref{Eq:ThetaPrewet}), it follows directly that $\theta_p > \theta_W$. 

Let us now consider the system off coexistence, again invoking eq.~(\ref{Eq:Adsorb}) as the condition determining $h$. A graphical solution of eq.~(\ref{Eq:Adsorb}) is sketched in Fig.~\ref{LambdaL}b. As long as $\theta > 0$, both relevant zeros of $\Lambda(h) L(h)$ (representing the solutions of eq.~(\ref{Eq:Adsorb}) for $H =0$) lie well within the interval $\left[h_-,h_+\right]$. For $H > 0$, the closed circle indicates again the stable solution. Obviously, the two points of intersection will merge when the dashed and solid curves touch each other only in a single point. This occurs at a certain curvature $H_p(\theta)$ of the liquid surface. For $H > H_p$, solid and dashed curve meet only for $h \rightarrow h_-$: there is no liquid adsorbed, and the substrate is dry. Hence the average position of the liquid surface, $h$, jumps discontinuously at $H = H_p$. It is clear from the construction that $H_p$ decreases monotonely with $\theta$. 

\begin{figure}[h]
\includegraphics[width = 7.5cm]{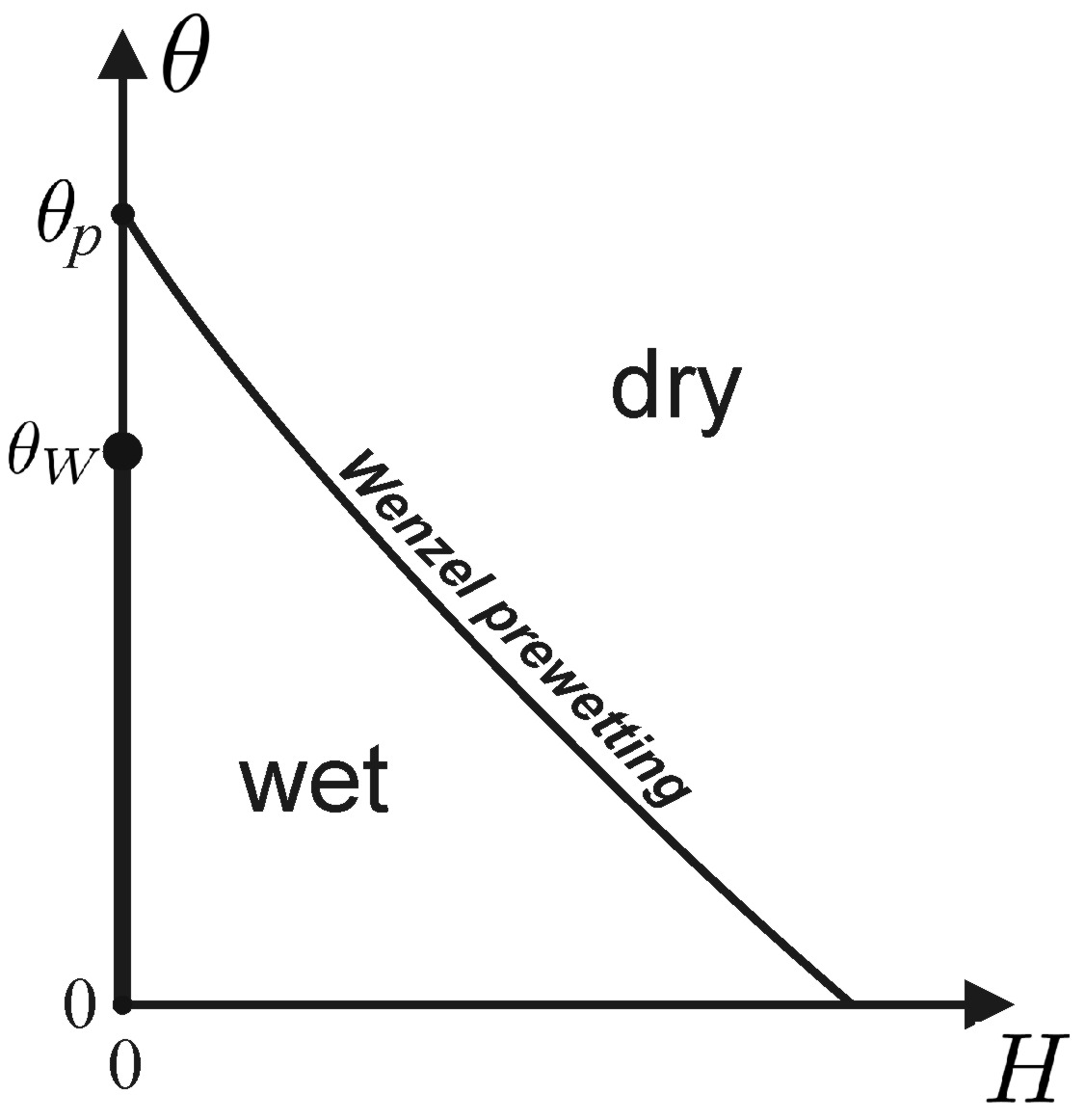}
\caption{The wetting phase diagram wetting on homogeneous, isotropically  rough surfaces. The most prominent feature is the existence of a 'Wenzel prewetting line' ($H_p(\theta)$), at which the adsorbed average film thickness jumps discontinuously from zero to a finite value.} 
\label{PhaseDiagramNonGauss}
\end{figure}

As $H$ is reduced below $H_p$,  $h$ increases continuously until at coexistence it reaches a value corresponding to the right zero of $\Lambda L$. Because of the phenomenological similarity of the jump in adsorbed film thickness to the prewetting transition encountered in standard wetting scenarios on flat substrates \cite{DietrichReview}, it has been  proposed to term this transition 'Wenzel prewetting' \cite{EPJE}. When the microscopic contact angle is varied, a prewetting line results, which is shown in Fig.~\ref{PhaseDiagramNonGauss} as the solid curve. As in the usual prewetting scenario, this line ends in a critical end point, when the solid and dashed curves in Fig.~\ref{LambdaL} intersect in only a single point. It is readily appreciated from the construction sketched in Fig.~\ref{LambdaL}, however, that this can occur only for $\theta \le 0$, which lies outside the physically accessible parameter range.

\begin{figure}[h]
\includegraphics[width = 8.5cm]{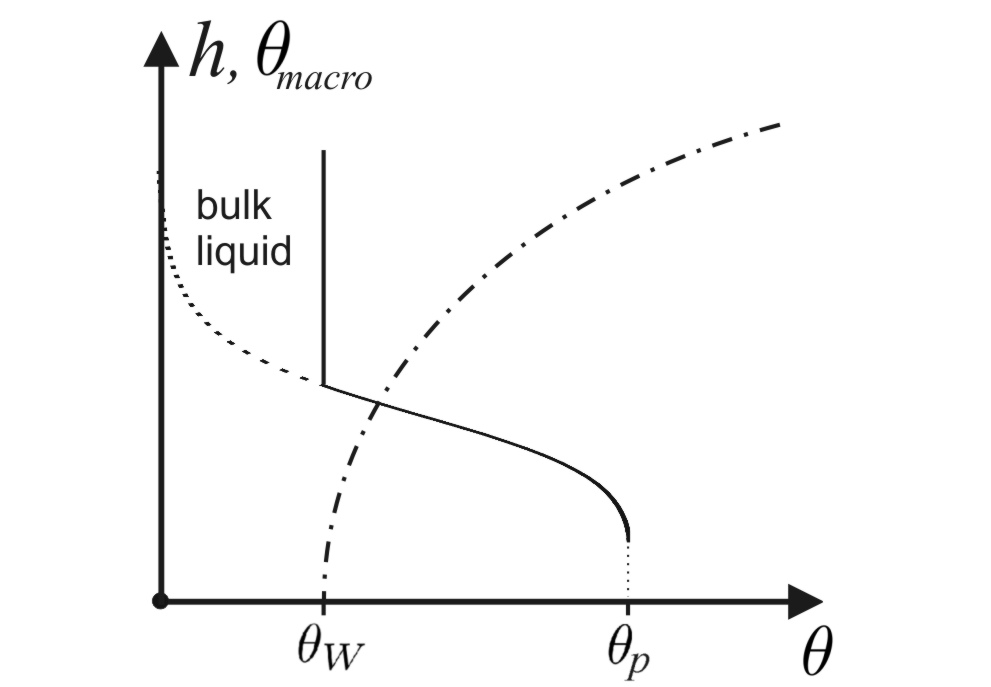}
\caption{Position of the liquid surface (solid curve) and macroscopic contact angle (dash-dotted curve) as a function of the microscopic contact angle at liquid-vapour coexistence. As $\theta$ comes below $\theta_p$, a macroscopic amount of liquid accumulates in the troughs and valleys of the roughness.  At $\theta \leq \theta_W$, the liquid surface detaches completely from the substrate, forming a bulk liquid layer of arbitrary thickness. The sigmiod solid curve corresponds to the right wing of $\Lambda (h)$, displayed in Fig.~\ref{LambdaL}a. \label{h(theta)} }
\end{figure}

The variation of $h$ along liquid/vapour coexistence, as $\theta$ is gradually decreased, can be directly read off Fig.~\ref{LambdaL}a, by inverting the right wing of $\Lambda(h)$. This is sketched in Fig.~\ref{h(theta)} as the solid curve which extends between $\theta_W$ and $\theta_p$, and continues in dashed below $\theta_W$. The liquid film first appears through a discontinuous jump at $\theta_p$ and increases gradually as $\theta_W$ is approached. As $\theta < \theta_W$, the liquid surface configuration which is bound to the surface topography through eq.~(\ref{Eq:YoungApprox}) becomes metastable (dashed curve in Fig.~\ref{h(theta)}), and the global minimum of the total free energy corresponds to the 'detached' liquid surface, or bulk liquid adsorption (vertical bold line in Fig.~\ref{PhaseDiagramNonGauss}).

It is interesting to discuss the expected shape of adsorption isotherms, i.e., to consider $h(H)$ at constant $\theta$. As it is obvious from Fig.~\ref{LambdaL}b, $h$ will remain finite for all positive $H$, reaching the value indicated in Fig.~\ref{h(theta)} by the sigmoid curve (solid and dashed) for $\theta < \theta_p$ and $H \rightarrow 0$.  This is remarkable, since there is complete wetting all along the coexistence line for $\theta < \theta_W$, such that for standard wetting scenarios, one would expect the adsorption isotherm to diverge continuously as $H \rightarrow 0$ \cite{DietrichReview}. In the present setting, this is only the case if $\theta = 0$. 

Let us finally get back to the macroscopic contact angle, $\theta_{macro}$. If $F(\theta)$ is the free energy per unit area of the sample for $\theta \ge \theta_W$, force 
balance at the macroscopic contact line yields 
\begin{equation}
\cos \theta_{macro} = \frac{F(\theta) - {\rm r}\gamma_{sl}}{\gamma}
\label{Eq:ThetaMacro}
\end{equation}
Since $F$ varies continuously with $\theta$, there will be a simple zero of $1-\cos\theta_{macro}$ at $\theta = \theta_W$. Hence $\theta_{macro}$ vanishes at $\theta_W$ as $\sqrt{\theta - \theta_W}$, in close analogy to first order wetting \cite{DietrichReview} (dash-dotted curve in Fig.~\ref{h(theta)}). At $\theta = \theta_p$, where $h$ jumps discontinuously to $h_p$, one might at first glance expect a jump as well in $\theta_{macro}$. However, since at $\theta_p$ the dry substrate is in coexistence with the patchy film with 'thickness' $h=h_p$, the free energies of the dry surface and the wetted surface must be equal, such that $F(\theta)$ is continuous even here. Consequently, there will be no jump in $\theta_{macro}$ at $\theta_p$. 

We should not close without mentioning that the presence of roughness gives rise to substantial equilibration times, as saddle points and extrema occurring at elevations $h_0 \in \left[h_-,h_+\right]$ provide effective pinning centers \cite{Joanny1984,DeGennes1985,Isich1992}. This will also result in contact angle hysteresis for $\theta_{macro}$. However, transport through either the gas phase or through the molecular adsorbed film \cite{Ala1996,Seemann2001} will always allow equilibration  over manageable times, since the lateral distances involved are never larger than the lateral length scale of the roughness.

Inspiring discussions with Daniel Tartakovsky, Siegfried Dietrich, Martin Brinkmann, J\"urgen Vollmer, Sabine Klapp, and Daniela Fliegner are gratefully acknowledged. The author furthermore acknowledges generous support form BP International Inc. within the ExploRe research program.

\end{document}